\documentstyle[psfig]{lamuphys}
\makeatletter
\let\chapter\hid@chapter
\makeatother

\begin{document}
\pagenumbering{arabic}

\title{$^{26}$\hspace{-0.2em}Al in the local
       interstellar medium}

\author{
J. Kn\"odlseder\inst{1},
K. Bennett\inst{5},
H. Bloemen\inst{3},
R. Diehl\inst{2},
W. Hermsen\inst{3},
U. Oberlack\inst{2},
J. Ryan\inst{4},
V. Sch\"onfelder\inst{2}, \and
P. von Ballmoos\inst{1}
}

\institute{
Centre d'Etude Spatiale des Rayonnements (CNRS/UPS), BP 4346, 31028
Toulouse Cedex, France
\and
Max--Planck--Institut f\"ur extraterrestrische Physik,
85740 Garching, Germany
\and
SRON-Utrecht, Sorbonnelaan 2, 3584 CA Utrecht, The Netherlands
\and
Space Science Center, University of New Hampshire, Durham NH 03824, U.S.A.
\and
Astrophysics Division, ESTEC, ESA, 2200 AG Noordwijk, The Netherlands
}

\authorrunning{J. Kn\"odlseder, K. Bennett, H. Bloemen, et~al.}

\maketitle

%
%
\newcommand{\al}{\mbox{$^{26}$\hspace{-0.2em}Al}}            
\newcommand{\na}{\mbox{$^{24}$\hspace{-0.1em}Na}}            
\newcommand{\Al}{\mbox{$^{27}$\hspace{-0.2em}Al}}            
\newcommand{\mg}{\mbox{$^{24}$\hspace{-0.1em}Mg}}            
\newcommand{\ti}{\mbox{$^{44}$\hspace{-0.1em}Ti}}            
\newcommand{\co}{\mbox{$^{56}$\hspace{-0.1em}Co}}            
\newcommand{\Done}{\mbox{D$_1$}}                             
\newcommand{\Dtwo}{\mbox{D$_2$}}                             
\newcommand{\Eone}{\mbox{E$_1$}}                             
\newcommand{\Etwo}{\mbox{E$_2$}}                             
\newcommand{\Etot}{\mbox{E$_{\rm tot}$}}                     
\newcommand{\phibar}{\mbox{$\bar{\varphi}$}}                 
\newcommand{\scdir}{\mbox{$(\chi,\psi)$}}                    
\newcommand{\de}{{\rm d}}                                    
\newcommand{\chisqr}{\mbox{$\chi^2$}}                        
\newcommand{\ei}{\mbox{$\{e_i\}$}}                           
\newcommand{\MeV}{\mbox{Me\hspace{-0.1em}V}}                 
\newcommand{\keV}{\mbox{ke\hspace{-0.1em}V}}                 
\newcommand{\Msol}{\hbox{M$_{\odot}$}}                       
\newcommand{\Rsol}{\hbox{R$_{\odot}$}}                       
\newcommand{\DEG}{\hbox{$^\circ$}}                           
\newcommand{\funit}{\mbox{ph cm$^{-2}$ s$^{-1}$}}            
\newcommand{\fster}{\mbox{ph cm$^{-2}$ s$^{-1}$ sr$^{-1}$}}  
\newcommand{\gray}{\mbox{$\gamma$-ray}}                      
\newcommand{\bgm}{background model}                          
\newcommand{\Mi}{\mbox{M$_{\rm i}$}}                         
\newcommand{\HI}{\mbox{H\hspace{0.2em}{\scriptsize I}}}      
\newcommand{\HII}{\mbox{H\hspace{0.2em}{\scriptsize II}}}    
\newcommand{\etal}{\mbox{{et~al.}}}                          
\renewcommand{\AA}{A\&A}                                     
\newcommand{\AAS}{A\&AS}                                     
\newcommand{\ApJ}{ApJ}                                       
\newcommand{\ApJS}{ApJ Suppl.}                               
\newcommand{\AJ}{AJ}                                         

\begin{abstract}

We estimate the 1.8 \MeV\ luminosity of the Sco-Cen association due
to radioactive decay of \al\ to $(4-15)\,10^{-5}$\funit.
We propose a low surface brightness, limb brightened bubble for the
1.8 \MeV\ intensity distribution.
The detectibility of this distribution with existing \gray\
telescopes is discussed.

\end{abstract}

\section{Introduction}

Gamma-ray line astronomy is a young, promising discipline which is
on the way to become a powerful diagnostic tool of nuclear astrophysics.
It allows the unambiguous identification of isotopic species in the
interstellar medium (ISM) by their spectral fingerprints: the characteristic
nuclear de-excitation lines.
The history of recent 
Galactic nucleosynthesis activity can be
studied by measurements of the 1.809 \MeV\ line arising from the decay
of radioactive \al.
Possible sources of \al\ are
core collapse supernovae (SNe), metal rich
novae, and massive stars with strong stellar winds (see review by Prantzos
\& Diehl 1996).
The distribution of \al\ was mapped by the \gray\ telescope
COMPTEL aboard the Compton Gamma-Ray Observatory (CGRO).
The 1.8 \MeV\ all-sky map (\cite{oberlack96}) clearly shows that almost
all emission is concentrated in the Galactic plane.
The irregular structure of the COMPTEL image with some
intermediate-latitude
features, and the appearance of distinct emission in the nearby Vela
and Cygnus regions, had led to speculations about a source
distribution, where a global Galactic nucleosynthesis glow underlies
emission from relatively few localized source regions with
particular recent nucleosynthesis activity (\cite{diehl96};
\cite{oberlack96}).
Local \al\ sources had been proposed before: Morfill \& Hartquist
(1985) suggested a SN event in the solar vicinity, Blake \&
Dearborn (1989) proposed SNe in the  Sco-Cen association --
the OB association nearest to the Sun -- as possible origin of the observed
\al.
The purpose of this paper is to revisit the \al\ contribution from
Sco-Cen based on recent nucleosynthesis calculations and new observational
constraints on the Sco-Cen history, and in view of the latest COMPTEL 1.809
\MeV\ measurements.

\section{Loop I}

Berkhuijsen \etal\ (1971) summarize observational evidence that Loop I,
a giant radio continuum loop centered on Sco-Cen, was created by supernova
explosions in the association.
Fejes \& Wesselius (1973) observe a \HI\ shell surrounding the radio continuum
loop some 5\DEG-15\DEG\ outside the best-fitting small circle.
Using ROSAT X-ray data, Egger (1993) developed a more detailed
scenario.
He claims that Loop I (the radiostructure and surrounding \HI\ shell) is
a superbubble
formed by stellar winds and SN explosions of the stars in Sco-Cen;
a recent supernova ($2\,10^5$yr ago) within the superbubble may have
re-heated
the gas, leading to the observed X-ray emission.

Based on Egger's model we estimate the \al\ production of Sco-Cen from
two
components: \al\ from the recent {\em re-heating} supernova and \al\
from the {\em older} supernovae (and Wolf-Rayet stars) which formed the
Loop I superbubble.
The \al\ yield of the recent SN can be estimated from the
earliest spectral type B0V in the association which corresponds
to an initial progenitor mass of 15-20\Msol.
According to nucleosynthesis calculations for type II SN about
$(3-9)\,10^{-5}$ \Msol\ of \al\ is expected for such a star
(\cite{timmes}).
The \al\ production of the events which formed the superbubble is estimated
by means of an analytic OB association evolution model which
predicts the \al\ output of an association as function of the association
age.
Stars in the mass interval 10-40 \Msol\ explode as type II SN and release
\al\ into the ISM at the end of their life.
Stars more massive than 40\Msol\ are assumed to exhibit a Wolf-Rayet (W-R)
phase during which they eject \al\ into the ISM by stellar winds.
Stellar and W-R lifetimes were taken from Schaller \etal\ (1992).
\al\ yields for W-R stars as function of initial stellar mass were
taken from Meynet \etal\ (1997), type II SN yields from Timmes \etal\
(1995).

\begin{figure}[thbp]
\psfig{file=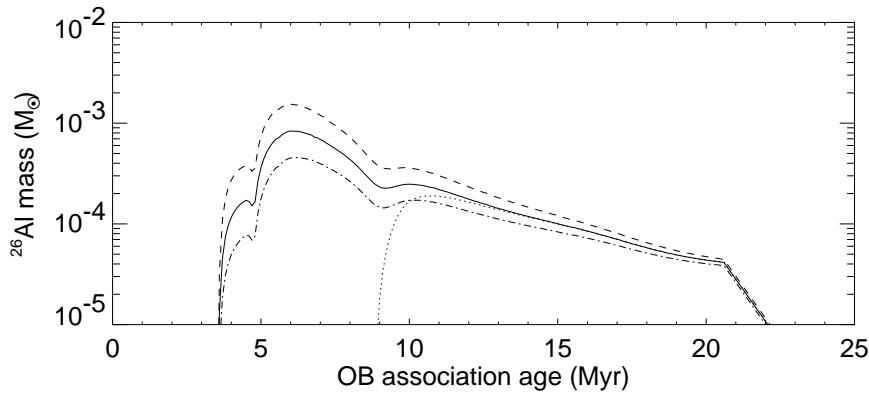,width=\hsize,angle=90,clip=}
\caption[]{\al\ lightcurves for the Sco-Cen OB association.
           Solid: $\Gamma=-1.5$, M$_{\rm up}=60$\Msol;
           dashed: $\Gamma=-1.0$, M$_{\rm up}=60$\Msol;
           dashed-dotted: $\Gamma=-2.0$, M$_{\rm up}=60$\Msol;
           dotted: $\Gamma=-1.5$, M$_{\rm up}=20$\Msol.}
\label{fig:evolve}
\end{figure}

The resulting \al\ lightcurve for Sco-Cen is shown in Fig.~\ref{fig:evolve}
for
different initial mass function (IMF) slopes $\Gamma$ and upper mass
limits M$_{\rm up}$.
The IMF was normalized to 42 stars with spectral type between B3
(7\Msol) and B1 (13\Msol) (\cite{bertiau}).
The general feature of the lightcurve is a short luminosity peak between
5 and 7 Myr after the formation of the assocation due to the explosion of
massive stars as supernovae.
The peak is preceeded by a small bump due to \al\ ejection by W-R stars
and followed
by a tail up to 21 Myr due to less massive supernova events.
After 21 Myr all stars more massive than 10\Msol\ exploded as supernovae,
hence the supply of potential \al\ sources is exhausted -- \al\
decays exponentially ($\tau_{26}=1.04\,10^6$yr).
It is clear from Fig.~\ref{fig:evolve} that the actual age of the association
is the most crucial parameter in the \al\ yield estimate -- the slope of
the IMF being of minor importance.
Taking an IMF slope of $\Gamma=-1.5$ and the age of Sco-Cen between 10-20
Myr (as estimated from the most massive member, Antares) results in an
\al\ yield of $(4-20)\,10^{-5}$\Msol.
We also applied the OB evolution model to the data of de Geus (1992) who
determined membership and age for each of the three subgroups of Sco-Cen
seperately.
Combining the \al\ lightcurves of the different subgroups gives a today \al\
mass of $(5-10)\,10^{-5}$\Msol.
The processed material from the early SN explosions which created
the superbubble was probably swept up by subsequent explosions and
could now be in a shell on the wall of the superbubble.
The ejecta of the most recent event, however, are expected to fill
the bubble more homogenously
due to turbulent mixing in the remnant interior (\cite{tt91}).
This scenario can be translated into a characteristic signature in
the angular
distribution of 1.8 \MeV\ \gray s:\ a low surface brightness
bubble would be surrounded by a circular emission limb.
The 1.8 \MeV\ limb would dominate the intensity distribution since
\al\ is concentrated in a much smaller region on the sky.
Taking the distance to Sco-Cen of 170 pc as the center of the bubble,
a bubble radius of 160 pc and a shell thickness of 10 pc yields
expected 1.8 \MeV\ fluxes of
$(3-11)\,10^{-5}$\funit\ and $(1-4)\,10^{-5}$\funit\ for the shell
and the bubble component, respectively.

The above scenario, however, is based on spherically symmetric SNR.
The ROSAT X-ray image of the Vela SNR shows
deviations from a spherical shell, which have recently been
interpreted
as high-velocity supernova ejecta (\cite{aschenbach}).
If Loop I obeys a similar morphology,
presence of supernova-generated \al\ far outside the classical
Loop I boundary would be possible.

\section{Observations}

The observation of diffuse low-intensity 1.8 \MeV\ emission as expected
from Loop I is difficult with COMPTEL.
Simulations show a tendency of our imaging techniques to translate
extended low-intensity emission into spot-like image noise (\cite{spie}).
Consequently weak emission features in the COMPTEL all-sky map at
medium and high Galactic latitudes could be an indication of local
\al\ (\cite{stmalo}).
The question on the presence of a diffuse \al\ component could be
addressed by combining observations of various telescopes:
large FOV instruments like SMM or GRIS give complementary information since
they are more sensitive to diffuse low intensity emission (\cite{diehl97}).
Indeed, the total 1.8 \MeV\ flux from the general direction of the Galactic
Center is higher
for these two instruments then the flux obtained with COMPTEL for the
Galactic plane, which indicates that a diffuse component is possibly missed
in the current COMPTEL analysis (\cite{diehl97}).
The flux discrepancy may be resolved with our \al\ predictions for Loop I.
Improving COMPTEL's sensitivity to diffuse
emission, and simultaneously analyzing data of COMPTEL, SMM and GRIS are
in progress.


\section*{Acknowledgments}
J. Kn\"odlseder is supported by the European Community through grant number
ERBFMBICT 950387.
The COMPTEL project is supported by the German government through
DARA grant 50 QV 90968, by NASA under contract NAS5-26645, and by
the Netherlands Organisation for Scientific Research NWO.


\end{document}